\documentclass[useAMS,usenatbib]{mn2e}

\usepackage[activeacute,swedish,english]{babel}
\usepackage{graphicx}
\usepackage[T1]{fontenc} 
\usepackage{latexsym}
\usepackage{verbatim}
\usepackage{amsmath}
\usepackage{subfigure}

\def\gtappeq{\mathrel{ \rlap{\raise.5ex\hbox{$>$}}
                      {\lower.5ex\hbox{$\sim$}}  } }

\def\leappeq{\mathrel{ \rlap{\raise.5ex\hbox{$<$}}
                      {\lower.5ex\hbox{$\sim$}}  } }

\title[The orbital period and system parameters of the recurrent nova
  T Pyx]{The orbital period and system parameters of the recurrent
  nova T Pyx} 
\author[H. Uthas, C. Knigge and D. Steeghs]{Helena Uthas$^{1}$\thanks{E-mail:
h.uthas@astro.soton.ac.uk}, Christian Knigge  $^{1}$ and Danny Steeghs$^{2}$\\\
$^{1}$University of Southampton, Department of Physics and Astronomy,
  Highfield, Southampton SO17 1BJ, UK\\ 
$^{2}$University of Warwick, Department of Physics, Coventry CV4 7AL, UK}

\begin{document}

\date{Accepted 2010 May 19.  Received 2010 May 18; in original form 2010 April 19} 

\pagerange{\pageref{firstpage}--\pageref{lastpage}} \pubyear{2010}

\maketitle

\label{firstpage}

\begin{abstract}

T Pyx is a luminous recurrent nova that accretes at a much higher rate
than is expected for its photometrically determined orbital period of
about 1.8 hours. We here provide the first spectroscopic confirmation
of the orbital period, $P =1.8295$ hours ($f = 13.118368 \pm 1.1
\times 10^{-5}$ c/d), based on time-resolved optical spectroscopy
obtained at the VLT and the Magellan telescopes. We also derive an
upper limit of the velocity semi-amplitude of the white dwarf,  $K1 =
17.9 \pm 1.6 \,$km$\,$s$^{-1}$, and estimate a mass ratio of $q = 0.20
\pm 0.03$. If the mass of the donor star is estimated using the
period-density relation and theoretical main-sequence mass-radius
relation for a slightly inflated donor star, we find $M_{2} = 0.14 \pm
0.03 \,$ M$_{\odot}$. This implies a mass of the primary white dwarf
of $M_{1} = 0.7 \pm 0.2 \,$ M$_{\odot}$. If the white-dwarf mass is
$>$ 1 M$_{\odot}$, as classical nova models imply, the donor mass must
be even higher. We therefore rule out the possibility that T Pyx has
evolved beyond the period minimum for cataclysmic variables. We find
that the system inclination is constrained to be $i \approx 10$
degrees, confirming the expectation that T Pyx is a low-inclination
system. We also discuss some of the evolutionary implications of the emerging physical picture of T Pyx. In particular, we show that epochs of enhanced mass transfer (like the present) may accelerate or even dominate the overall evolution of the system, even if they are relatively short-lived. We also point out that such phases may be relevant to the evolution of cataclysmic variables more generally.

\end{abstract}


\section{Introduction}

Binary systems containing a Roche-lobe filling main-sequence star that
loses mass to a primary white dwarf (WD) are referred to as cataclysmic
variables (CVs, see~\citealp{b28} for an overview). T Pyx is a
luminous CV in which the donor star is transferring mass at a very 
high rate onto a high-mass white dwarf, resulting in unusually frequent
thermo-nuclear runaways (TNRs) on the surface of the 
primary. Between the years 1890 $-$ 1967, T Pyx has undergone five
such nova eruptions, with an recurrence time of about 20 years between the eruptions, and was therefore classified as a member of the recurrent nova (RN) subclass. However, the last eruption was in 1966, which means that T Pyx has now passed its mean
recurrence time by more than 20 years. The eruptive behavior of RNe in
comparison with classical novae is thought to be due to a high
mass-transfer rate in combination with a massive primary white dwarf
~\citep{b31}.  

~\cite{b13} carried out an extensive photometric study of
T Pyx and found a stable, periodic signal at $P=1.83$ hours that was
interpreted as the likely orbital period. This would place T Pyx below
the CV period gap and suggests a donor mass around $M_{2} \sim 0.1$
M$_{\odot}$. This is surprising. According to standard evolutionary
models, a CV below the period gap should be faint and have a low
accretion rate driven primarily by gravitational radiation (GR). Yet T Pyx's quiescent luminosity and status as a RN both imply that it has a high accretion rate of $> 10^{-8}$ M$_{\odot}$ yr$^{-1}$ (\citealp{b13};~\citealp{b20}). This is about two orders of magnitude higher than expected for ordinary CVs at this period. 

Assuming that the determination of the photometric orbital period is correct, the
existence of T Pyx is interesting for at least two reasons. First, 
unless we are seeing the system in a transient evolutionary state, its
lifetime would have to be very short $\tau \sim M_{2}/\dot{M}_{2} \leappeq
10$~Myrs. This would imply the existence of an evolutionary channel
leading to the fast destruction of at least some short-period
CVs. Second, TNRs frequent enough to qualify as RNe are thought to be possible only on
high-mass accreting WDs ($M_{1} \gtappeq 1$ M$_{\odot}$). Moreover, RNe are
the only class of novae in which the WD is expected to gain more mass
between eruptions than it loses during them. This would make T Pyx a
strong candidate Type Ia supernova progenitor.  

However, the recent study of the system by~\cite{b18} (see also~\citealp{b20}) suggests,
first, that T Pyx {\em is}, in fact, in a transient evolutionary
state, and, second, that, integrated over many nova eruptions, its WD
does lose more mass than it gains. More specifically,~\cite{b18} suggest that T Pyx was an 
ordinary cataclysmic variable until it erupted as a nova in 1866. This
eruption triggered a wind-driven supersoft X-ray phase (as 
first suggested by~\citealp{b7}), resulting in an unusually high
luminosity and accretion rate. However, unlike in the original
scenario proposed by~\cite{b7}, the supersoft phase is not
self-sustaining, so that the accretion rate has been
declining ever since the 1866 nova eruption from $\dot{M} \sim 10^{-7}$
M$_{\odot}$\,yr$^{-1}$ to $10^{-8} $M$_{\odot}$\,yr$^{-1}$. As a result, T
Pyx has faded by almost 2 magnitudes since the nova eruption
(\citealp{b18}). Based on this, and the fact
that T Pyx has already passed its mean recurrence time by more than 20
years,~\cite{b18} argue that T Pyx might no
longer even be a recurrent nova. If these ideas are correct, T Pyx is
not a viable SN Ia progenitor, and its remaining lifetime can be
substantially longer than a few million years. However, if all its
ordinary nova eruptions are followed by relatively long-lived (> 100
yrs) intervals of wind-driven evolution at high $\dot{M}$, its secular
evolution may nevertheless be strongly affected, with significant
implications for CV evolution more generally (see also Section~\ref{SD}).

A key assumption in virtually all of these arguments is that the
photometric period  measured by~\cite{b13} is, in 
fact, the orbital period of the system. So far, there has only been
one attempt to obtain a spectroscopic period for T Pyx, by~\cite{b26},
who reported a spectroscopic modulation
with $P=3.44$ hours. Such a long orbital period above
the CV period gap would be much more consistent with the high
accretion rate found in T Pyx. In this study, we present the first
definitive spectroscopic determination of the orbital period of T Pyx,
showing 
that it is, in fact, consistent with Patterson et al.'s photometric
period. We also use our time-resolved spectroscopy to estimate the
main system parameters, such as the velocity semi-amplitude of the
white dwarf ($K1$), the mass-ratio (q), the masses ($M_{1}$ and
$M_{2}$) and the orbital inclination ($i$). Finally, we discuss the
implications of our results for the evolution of T Pyx and related
systems.

\section{Observations and Their Reduction}
 
 \subsection{VLT Multi-fibre Spectroscopy}
 
Multi-fibre Spectroscopy of T Pyx was obtained during five nights in
2004 and 2005 with the GIRAFFE/FLAMES instrument mounted on the Unit
Telescope 2 of the VLT at ESO Paranal, Chile. The data were taken in
the integrated field-unit mode. The total field of view in this mode
is about 11.5" $\times$ 7.3" and thus covers most of T Pyx's 10"
diameter nova shell (\citealp{b29}). We used the fibre system ARGUS,
which consists of 317 fibres distributed across the field, of which 5
are pointing to a calibration unit and 15 are pointing to sky. The
grating order was 4, which gives a resolution of R=12000. The
wavelength range was chosen between 4501 to 5078 \AA, so that the
emission-line spectrum would include the Bowen blend at 4645 -- 4650
\AA\, and the HeII at 4686 \AA. With this setup, the dispersion is
0.2 \AA/pix, corresponding to about 12.5 km\,s$^{-1}$/pix. The full
widths at half-maximum (FWHMs) of a few spectral lines obtained
simultaneously with the science spectra from fibres pointing to the
calibration unit indicate that the spectral resolution is about 0.4
\AA. A log of the observations can be found in Table~\ref{tab:obs}.

The initial steps in the data reduction were performed using the ESO
pipeline for GIRAFFE. The pipeline is based on the reduction software
BLDRS from the Observatory of Geneva. The basic functions of the
pipeline are to provide master calibration data and dispersion
solutions. The pipeline also provides an image of the reconstructed
field of view, which can be used to associate a specific fibre to a
given object. In order to extract the spectrum from the desired fibre
and to correct for the contribution from the sky background and cosmic
rays, the output from the pipeline was processed further in IRAF. The
PSF of our target, T Pyx, is spread out over several fibres and 6
single fibre spectra containing significant target flux were extracted
and median combined after weighting each spectrum by the mean flux in
the region 4660 - 4710 \AA (covering HeII at 4686 \AA). Cosmic rays
were removed by first binning the spectra over 7 pixels (the cosmic rays
have a typical width of 2 $-$ 6 pixels). The smoothed spectra were
then subtracted from the corresponding fibre spectra to only leave the
residuals and the cosmic rays. Cosmic rays were then removed and the
residuals added back to the target spectra. Fibres containing the sky
were extracted and combined to create a master-sky spectrum. Finally,
the master-sky was subtracted from the combined science spectrum.

\subsection{Magellan Long-slit Spectroscopy}

T Pyx was observed again during four nights in March 2008, this time
on the 6.5 meter Baade telescope Magellan I at Las Campanas, Chile
(see Table~\ref{tab:obs} for a log of the observations). Long-slit spectroscopy
was obtained with the instrument IMACS using the Gra-1200-17.45 grating
with a 0.9" slit, resulting in a dispersion of 0.386 \AA/pix,
corresponding to about 25 km\,s$^{-1}$/pix. We estimate that the spectral
resolution of these observations is about 1.5 \AA, as measured from
the FWHMs of a few spectral lines in the arc-lamp spectra. The overall
spectra span over four CCDs and cover a total wavelength range of 4000
-- 4800 \AA. All frames from the four CCDs were treated separately
during both the 2D and 1D reduction steps.
 
The spectra were reduced in IRAF using standard packages. A master
bias produced from combining all bias frames obtained during the four
nights was subtracted from the science frames and the overscan region
was used to remove the residual bias. A master flat-field, corrected
for illumination effects was produced. The spectra were then
flat-field corrected and extracted. Line identifications of the ThArNe
lamp spectra obtained during the nights were carried out with the help
of NOAO Spectral Atlas Central and the NIST Atomic Spectra
Database. The spectra were cosmic ray rejected using the method
described previously for the VLT dataset. A flux calibration was done
using data of the white-dwarf flux-standard star EG274, obtained
during the same nights as the target spectra. Also, extinction data
from the site and flux reference data of EG274 from ~\cite{b5} were used.  \\\\ Analysis of the reduced VLT and Magellan
data were carried out using the packages MOLLY and DOPPLER, provided by Tom Marsh.

\begin{table}
  \centering
  \begin{minipage}{70mm}
  \caption{Log of the observations (the date is according to UT time at the start of the night).}
 \begin{tabular}{llll}
  \hline
 Date & Tel/Inst. & No. of Exp. & Exp.[s] \\
 \hline
041210 & VLT/GIRAFFE & 28 & 180\\
041224 & VLT/GIRAFFE & 28 & 180\\
050127 & VLT/GIRAFFE & 29 & 180\\
050128 & VLT/GIRAFFE & 28 & 180\\
050131 & VLT/GIRAFFE & 29 & 180\\
080316 & MAGELLAN/IMACS & 37 & 240\\ 
080317 & MAGELLAN/IMACS & 62 & 240\\ 
080318 & MAGELLAN/IMACS & 62 & 240\\
080319 & MAGELLAN/IMACS & 39 & 240\\
\hline
\label{tab:obs} 
\end{tabular}
\end{minipage}
\end{table}

\section{Data Analysis}

\subsection{The Overall Spectrum}

\begin{figure*}
\centering
\subfigure[The normalized and averaged spectra constructed from 140 individual exposures obtained at the VLT telescope.] 
{
    \label{fig:spec_vlt}
    \includegraphics[width=8cm]{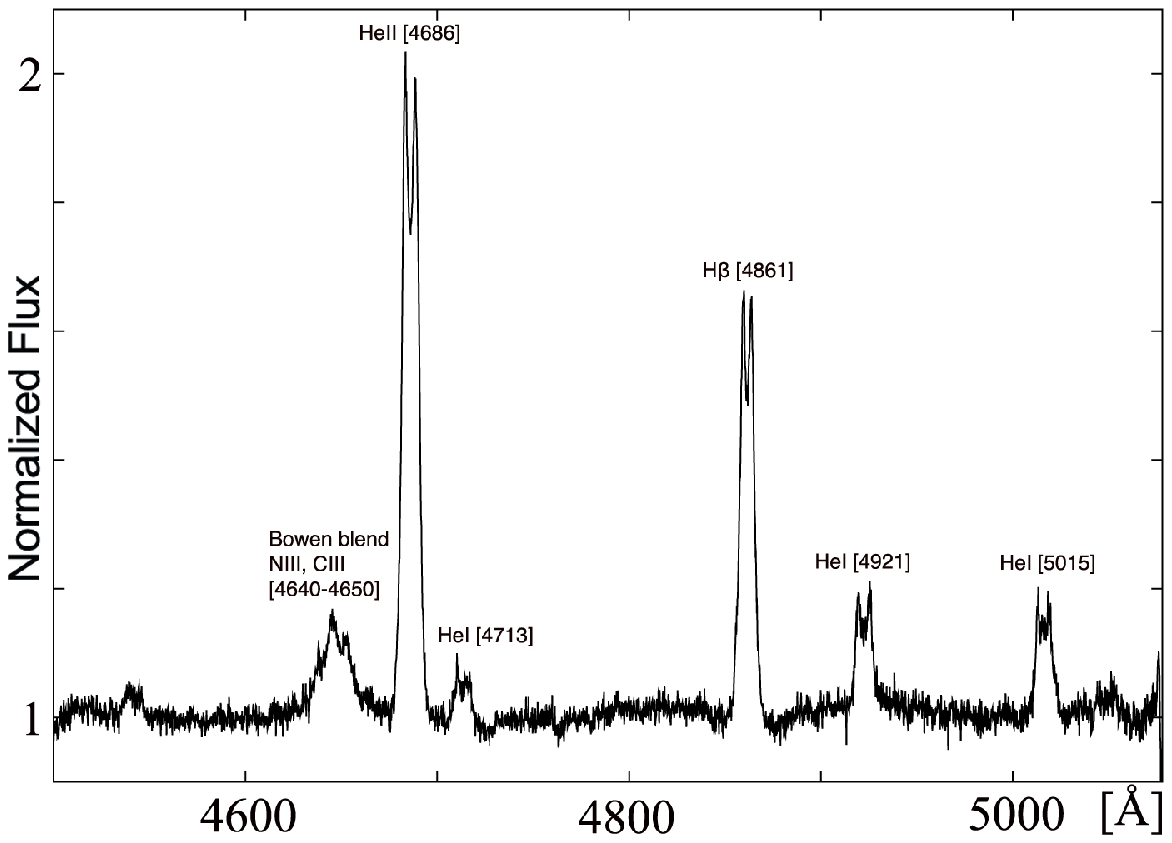}
}
\hspace{0.2cm}
\subfigure[The flux-calibrated average of 200 spectrum observed with the Magellan telescope.] 
{
    \label{fig:spec_mag}
    \includegraphics[width=8.5cm]{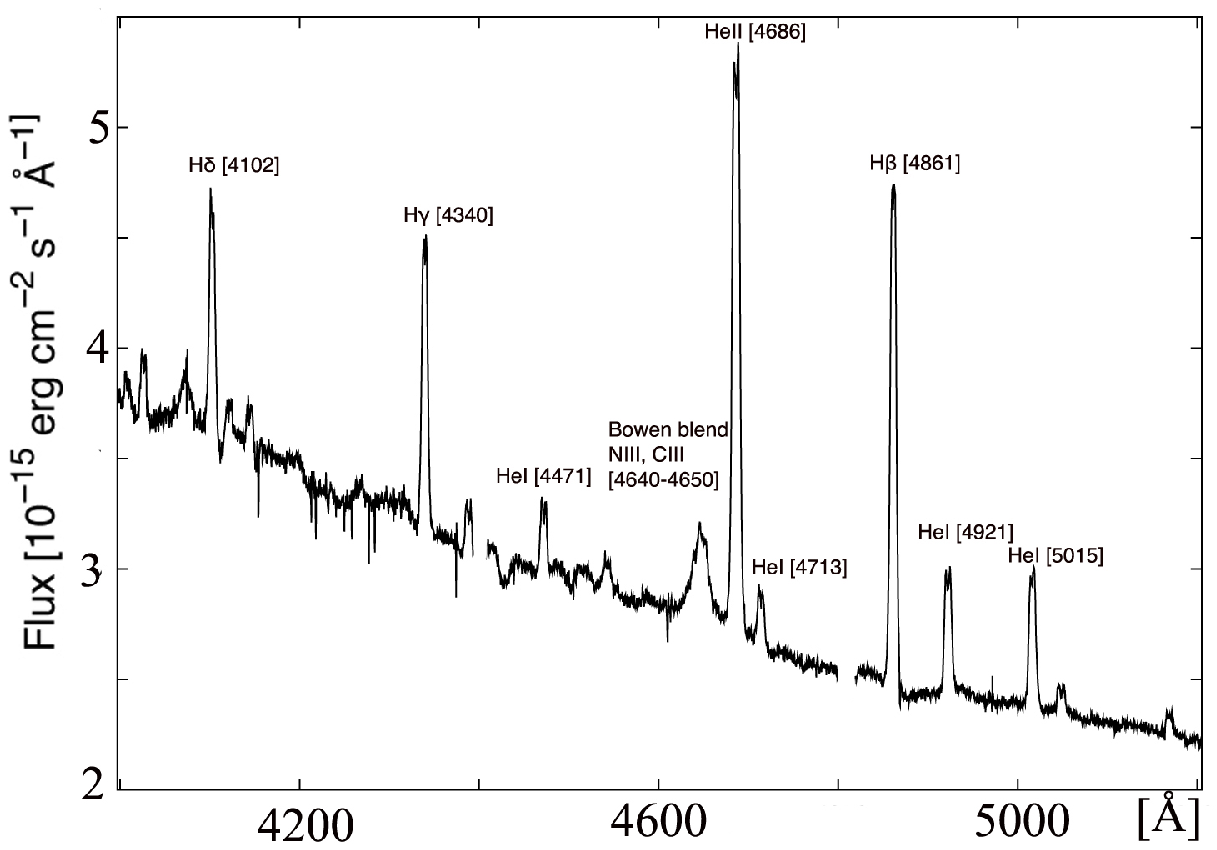}
}
\caption{}
\label{fig:sub} 
\end{figure*}

T Pyx has a high-excitation emission-line spectrum that is unusual
when compared to systems with a low $\dot{M}$, but the overall spectrum is similar to that for nova-like variables, supporting the idea that T Pyx has a high accretion rate. Its brightness has been found to be fading for the last century, indicating a decrease in $\dot{M}$,
and its 2009 magnitude in blue was B = 15.7 mag (\citealp{b18}). Double-peaked lines originating from the disc are seen in
the spectrum, and the bright disc outshines any spectral signature of
the primary WD and the donor star. The strongest feature is the
doubled-peaked HeII line at 4686 \AA. Double-peaked lines identified
as HeI at $\lambda$4713, $\lambda$4921 and $\lambda$5015, as well as
the Balmer lines, are also present (Figure~\ref{fig:spec_vlt} and Figure~\ref{fig:spec_mag}).

The Bowen blend at $\lambda$4640 $-$ $\lambda$4650 is clearly visible
and consists of several lines of NIII and CIII. The Bowen blend is
also seen in other CVs and in low-mass X-ray binaries with
high-accretion rates. In some cases, it has been related to emission
from the donor star (eg.~\citealp{b24}), where it is thought
to be produced by a fluorescence process as the front side of the
donor is strongly irradiated by the hot accretor. The Bowen blend was
clearly visible in the spectrum of T Pyx obtained by~\cite{b10}, and based on this, we were hoping to detect narrow components
associated with the donor. However, no narrow donor star features were
found in the blend.  

Data analysis was carried out for all the strongest lines but
our final results are based on analysis of the HeII line at 4686 \AA \, and the HeI line at 4921 \AA.


\begin{figure}
\includegraphics[width=9cm]{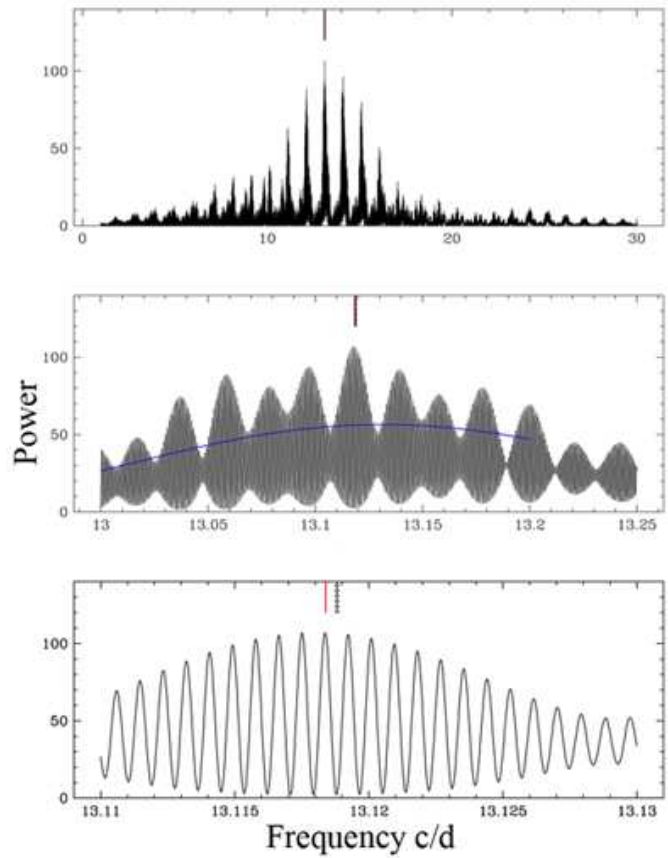}
\caption{
Power spectrum of the HeII radial-velocity data, obtained from the
combined VLT and Magellan data sets. The vertical grey line in the
bottom panel represents the best period found by Patterson et al. 1998, along with its error. The red line represents the recalculated period using the latest photometric ephemeris (see Section~\ref{PE}) where the period derivative has been taken into account (Patterson et al. in preparation).  }
\label{fig:pow}
\end{figure}


\subsection{The Orbital Period} 

\subsubsection{The Photometric Ephemeris} \label{PE}

The key goal of our study is to
obtain a definitive determination of T Pyx's orbital period based on 
time-resolved spectroscopy. More specifically, we wish to test if the
photometric modulation reported by~\cite{b13} is orbital
in nature. The photometric ephemeris is given by~\cite{b13} as
\\\\
Minimum light = HJD $2446439.428 + 0.0762233E + 3.5 \times 10^{-11}E^{2}$. 
\\\\
The quadratic term is highly significant and indicates a very high period derivative of $\dot{P}
\simeq 9 \times 10^{-10}$ (this corrects a typo in the original paper). The period is found to be increasing, which is not expected for a system that has not yet passed the minimum period. If the photometric signal is orbital, the current
evolutionary time-scale of the system is only $\tau \sim P / \dot{P}
\sim 3 \times 10^{5}$ yrs. Such a short period change time-scale is highly unusual for a CV.  

The Center for Backyard Astronomy (CBA) has continued to gather additional
timings in the period 1996 -- 2009. This new photometry, along 
with an updated ephemeris, will be published in due course (Patterson
et al., in preparation). However, since this new data set is much more
suited to a comparison against our spectroscopic observations (since it 
overlaps in time), Joe Patterson has kindly made this 
latest set of CBA timings available to us. Our own prelimary analysis
of all the existing timings yields a photometric ephemeris of
\\\\
Minimum light = HJD(UTC) $2451651.65255(35) + 0.076227249E(16) + 2.546(54) \times 10^{-11}E^{2}$.
\\\\

The numbers in parenthesis are the uncertainties on the two least significant digits, and were estimated via bootstrap simulations. Taking the period change into account, the
average period at the time of the optical spectroscopy was $P = 0.076228860 \pm 3.0 \times 10^{-8}$ d (corresponding to $f = 13.1183912 \pm 5.2 \times 10^{-6}$ c/d). 
\\\\
This new ephemeris confirms that the period continues to increase at a fast
rate.  More specifically, the current estimate of the period
derivative is  $\dot{P} = (6.68 \pm 0.14) \times 10^{-10}$, corresponding to a timescale of
$P/\dot{P} \simeq 3 \times 10^{5}$ yrs.  It is important to take this period
derivative into account when comparing spectroscopic and photometric
periods.

\subsubsection{The Spectroscopic Period}

In order to establish the orbital period spectroscopically, we carried
out a radial-velocity study using the double-Gaussian method~\citep{b19}. This method is most effective in the line wings, which are formed in the inner regions of the accretion disc. As
discussed further in Section~\ref{RV}, we tried a variety of Gaussian
FWHM's, as well as a range of separations between the two
Gaussians.

 The results shown in this section are for a FWHM =  450 km\,s$^{-1}$
and a separation of 300 km\,s$^{-1}$, and are based on the full combined VLT +
Magellan data set, spanning the time period December 2004 to March
2008. The radial-velocity study was carried out for several spectral lines but for the purpose of this section, we will focus exclusively on the radial-velocities derived from the strong HeII line at 4686 \AA, since this line provides the most accurate results. The power spectrum~\citep{b32} of this spectroscopic data set is presented in Figure~\ref{fig:pow}. The top panel, covering a broad frequency range, confirms that there is a strong signal near the photometrically
predicted 13 c/d. No signal is found at $P=3.44$ hours, corresponding to the period found by~\cite{b26}. The middle and bottom panels show close-ups of
narrower frequency ranges around this power spectral peak. Note, in
particular, that the red line in the bottom panel marks the
photometrically predicted frequency for the epoch corresponding to the
mid-point of our spectroscopic data set. For comparison, we also show
the frequency of the linear ephemeris determined by~\cite{b13} for just the 1996-1997 timings, along with its error. The spectroscopic and photometric periods appear to be mutually
consistent, {\em but only if we account for the photometry period derivative}. More specifically, the  
$\dot{P}$-corrected photometric prediction lies extremely close to the
peak of the most probable spectroscopic alias. 

This apparent consistency can be checked more quantitatively. The
orbital frequency predicted by the photometry for the average
epoch of our spectroscopy is  $f_{phot} = 13.1183912 \pm  5.2 \times
10^{-6}$~c/d, where the error is once again based on bootstrap
simulations.  Similarly, we have carried out a bootstrap error
analysis for the frequency determined 
from our spectroscopy. Since we are only interested in the agreement
between the photometry and the spectroscopy, we only considered the
spectroscopic alias closest to the photometrically 
determined frequency (this is also the highest alias). This yielded
$f_{spec} = 13.118368 \pm 1.1 \times 10^{-5}$~c/d. The difference
between these determinations formally amounts to just under
2$\sigma$. We consider this to be acceptable agreement. 

The level of agreement between spectroscopic and
photometric periods is important. The spectroscopy on its own is sufficient to establish beyond doubt that T Pyx is a short-period system with $P_{orb} \simeq 1.8$~hrs. However,
the fact that we can establish reasonable consistency with the
photometric data if (and only if) we account for the photometric
period derivative suggests that (i) we can trust the much higher
precision photometric data to provide us with the most accurate
estimate of the orbital period, and (ii) that the large period
derivative suggested by the photometry is, in fact, correct. 

\subsection{Trailed Spectrograms and Doppler Tomography}

\begin{figure*}
\centering
\subfigure[Magellan, HeII(4686)] 
{
    \label{fig:sub:b}
    \includegraphics[width=4cm]{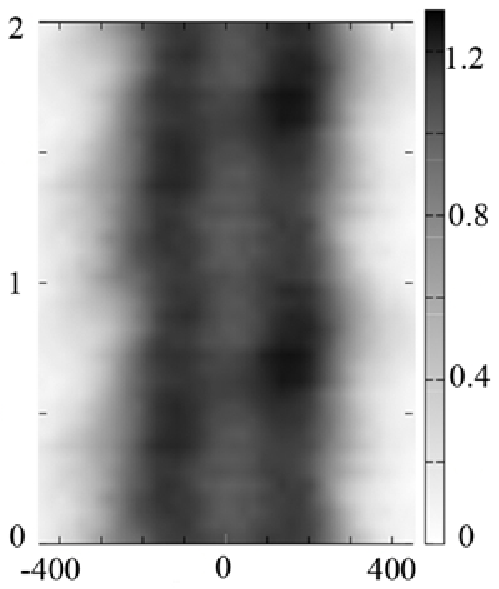}
}
\hspace{0.0cm}
\subfigure[Magellan, HeI(4921)] 
{
    \label{fig:sub:c}
    \includegraphics[width=4cm]{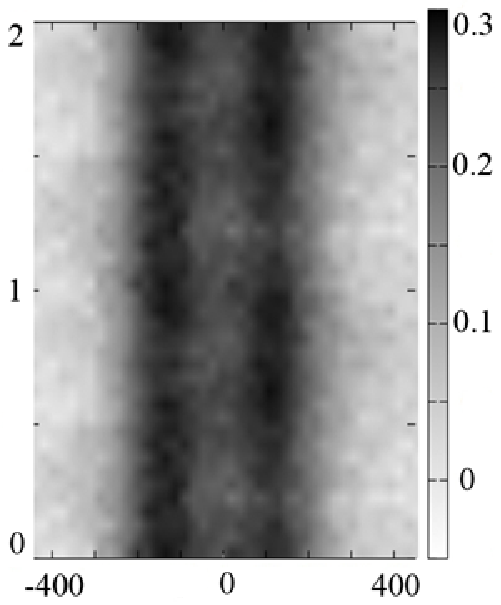}
}
\hspace{0.0cm}
\subfigure[VLT, HeII(4686)] 
{
    \label{fig:sub:a}
    \includegraphics[width=4.3cm]{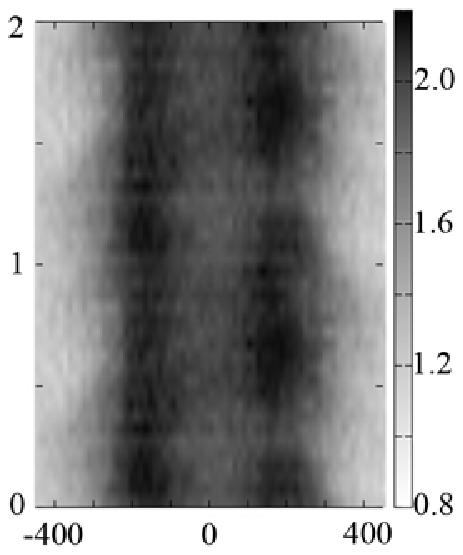}
}
\hspace{0.0cm}
\subfigure[VLT, H$\beta$(4861)] 
{
    \label{fig:sub:d}
    \includegraphics[width=4cm]{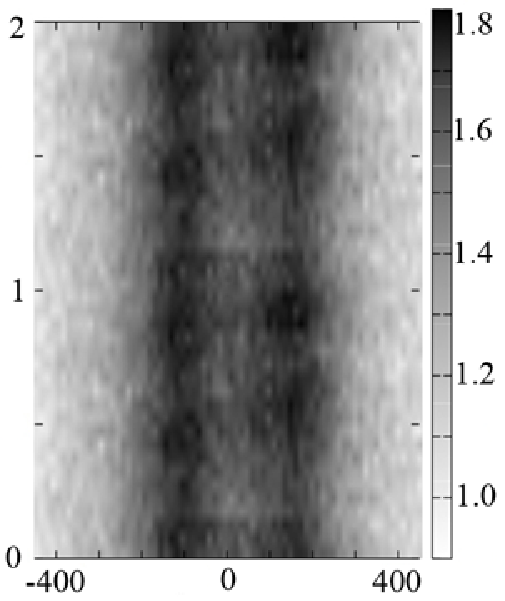}
}
\caption{Trailed spectrograms, phase binned to match the orbital
  resolution (37 bins for the VLT data and 27 bins for the Magellan
  data), are here plotted over 2 periods. Phases are plotted against velocities (in km\,s$^{-1}$).} 
\label{fig:spec} 
\end{figure*}

In order to visualize the orbital behavior of the spectral profiles, 
trailed spectrograms were constructed for HeII, HeI, H$\beta$ and the
Bowen blend. These spectrograms, shown in Figure~\ref{fig:spec}, are phase binned to
match the orbital resolution of our data (37 bins for the VLT data and
27 bins for the Magellan data), and are here plotted over 2
periods. All lines show emission features moving from the blue to the
red wing, and localized emission is seen in the red wing at phase, $\phi \approx
0.6$. Similar structures are seen in the spectrograms for both the
HeII and the Bowen blend indicating that they originate in the
same line-forming region, presumably the accretion disc.

\begin{figure*}
\centering
\subfigure[Magellan, HeII(4686)] 
{
    \label{fig:sub:a}
    \includegraphics[width=5.7cm]{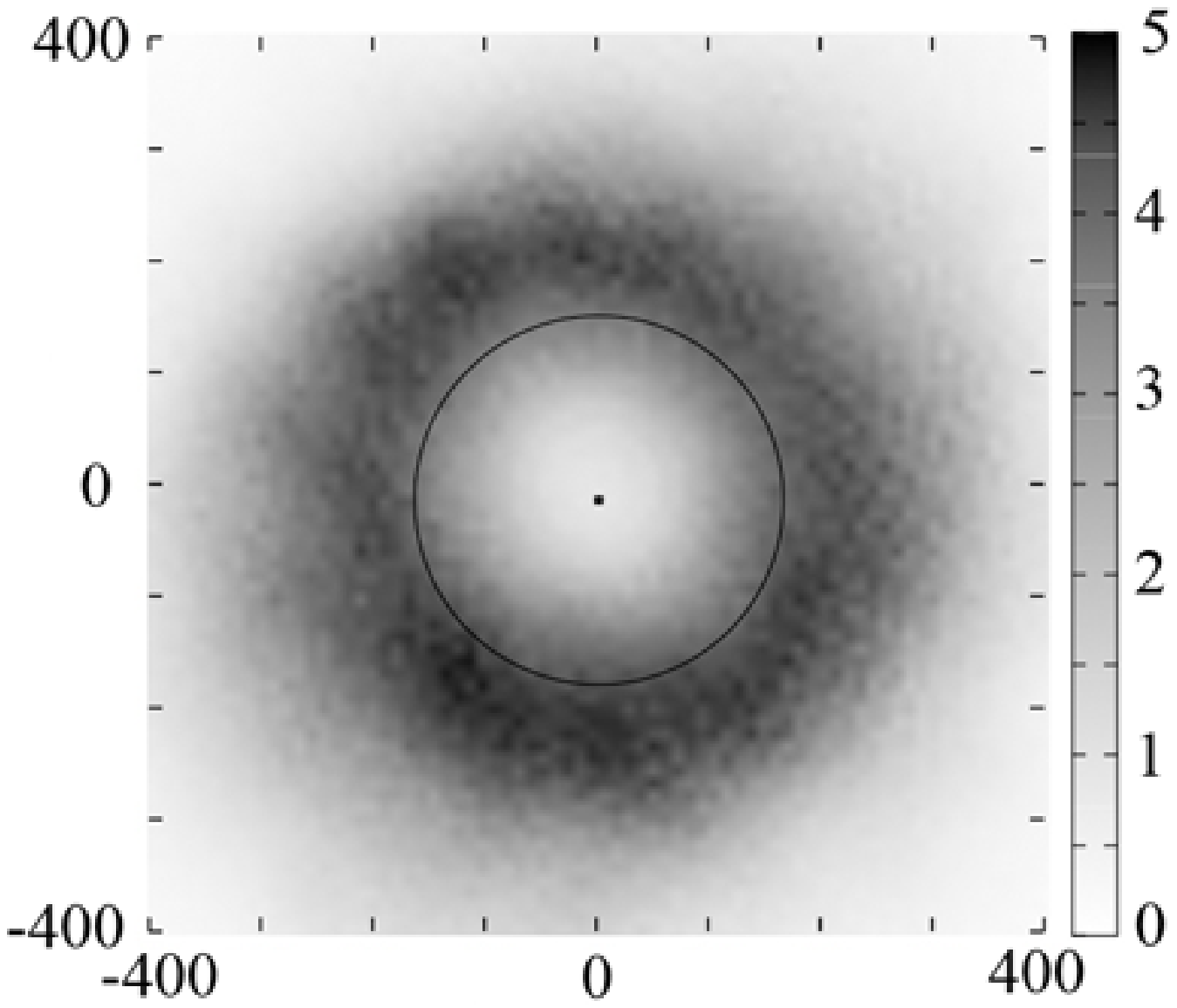}
}
\hspace{0.0cm}
\subfigure[Magellan, Bowen blend NIII(4641) and CIII(4650, 4647)] 
{
    \label{fig:sub:b}
    \includegraphics[width=5.5cm]{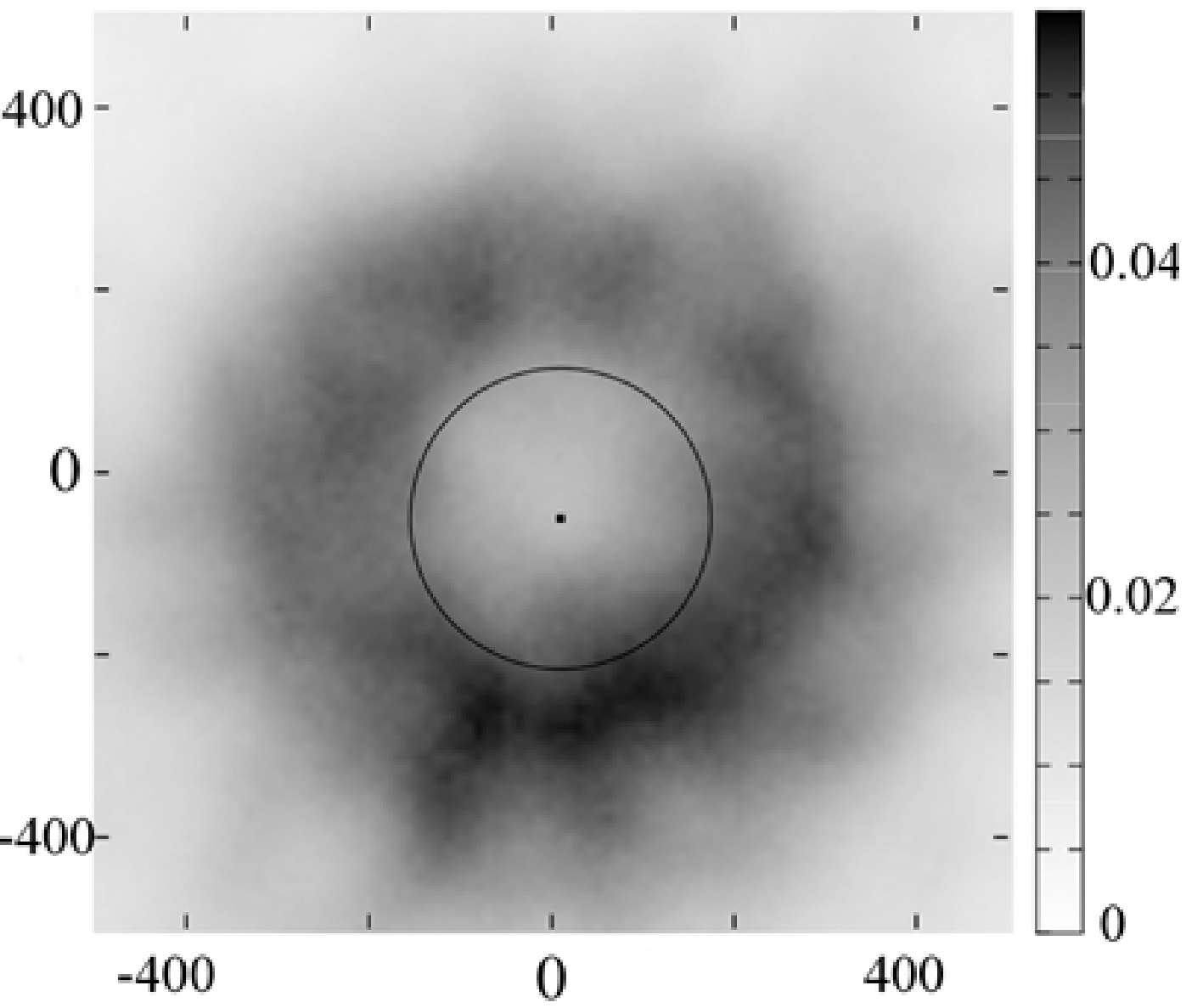}
}
\hspace{0.0cm}
\subfigure[Magellan, HeI(4921)] 
{
    \label{fig:sub:c}
    \includegraphics[width=5.5cm]{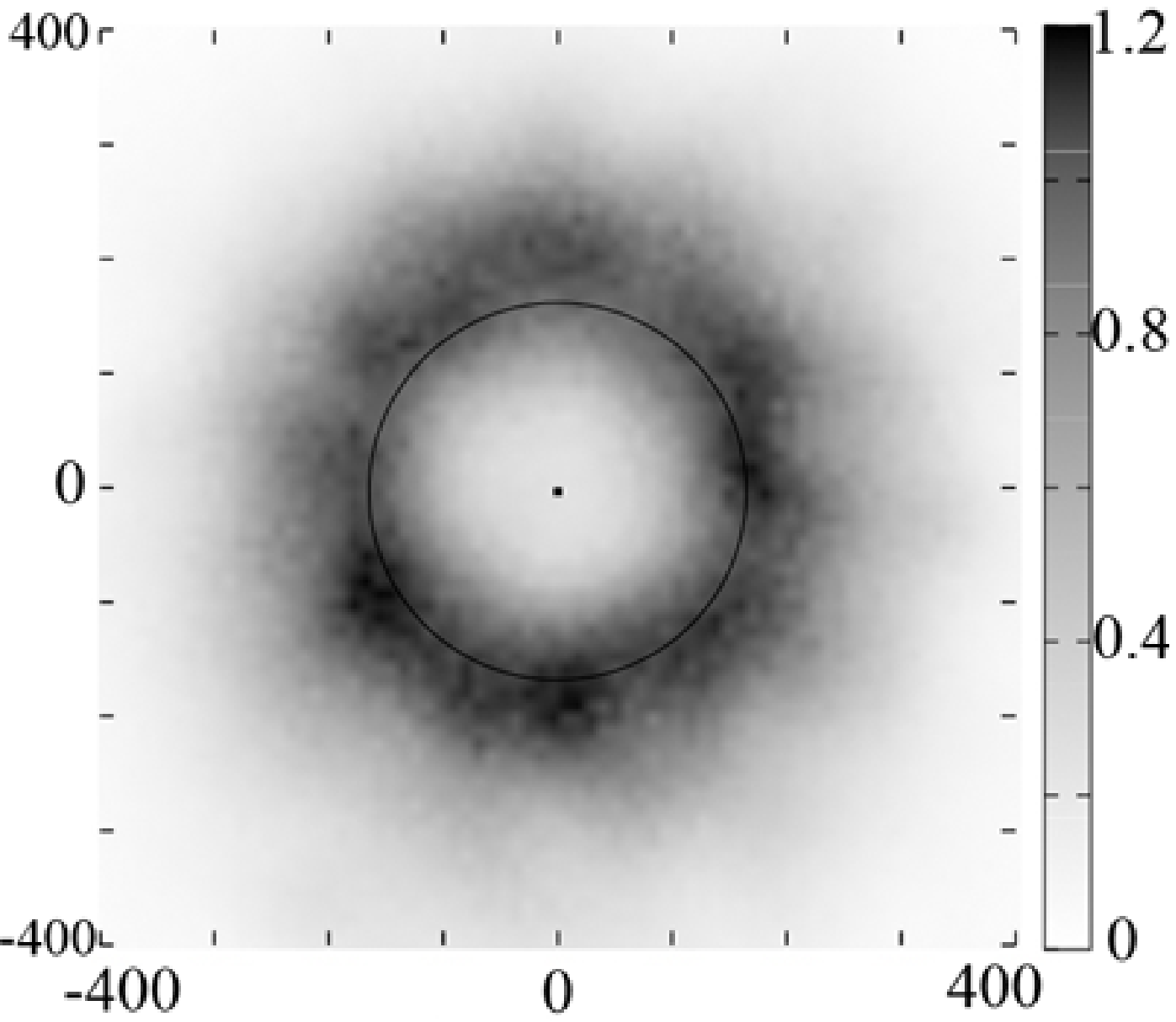}
}
\hspace{0.0cm}
\subfigure[VLT, HeII(4686)] 
{
    \label{fig:sub:d}
    \includegraphics[width=5.7cm]{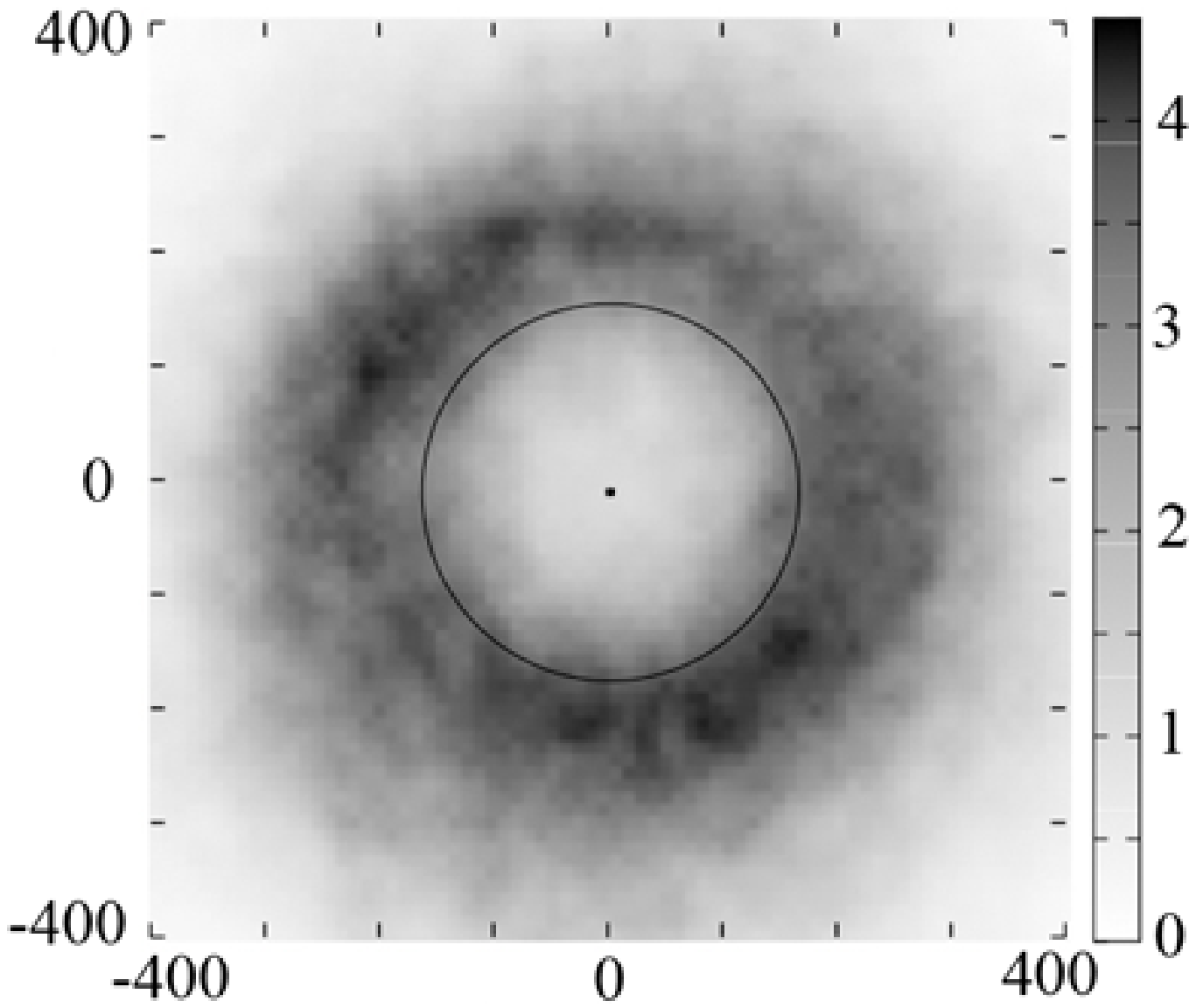}
}
\hspace{0.0cm}
\subfigure[VLT, Bowen blend NIII(4641) and CIII(4650, 4647)] 
{
    \label{fig:sub:d}
    \includegraphics[width=5.5cm]{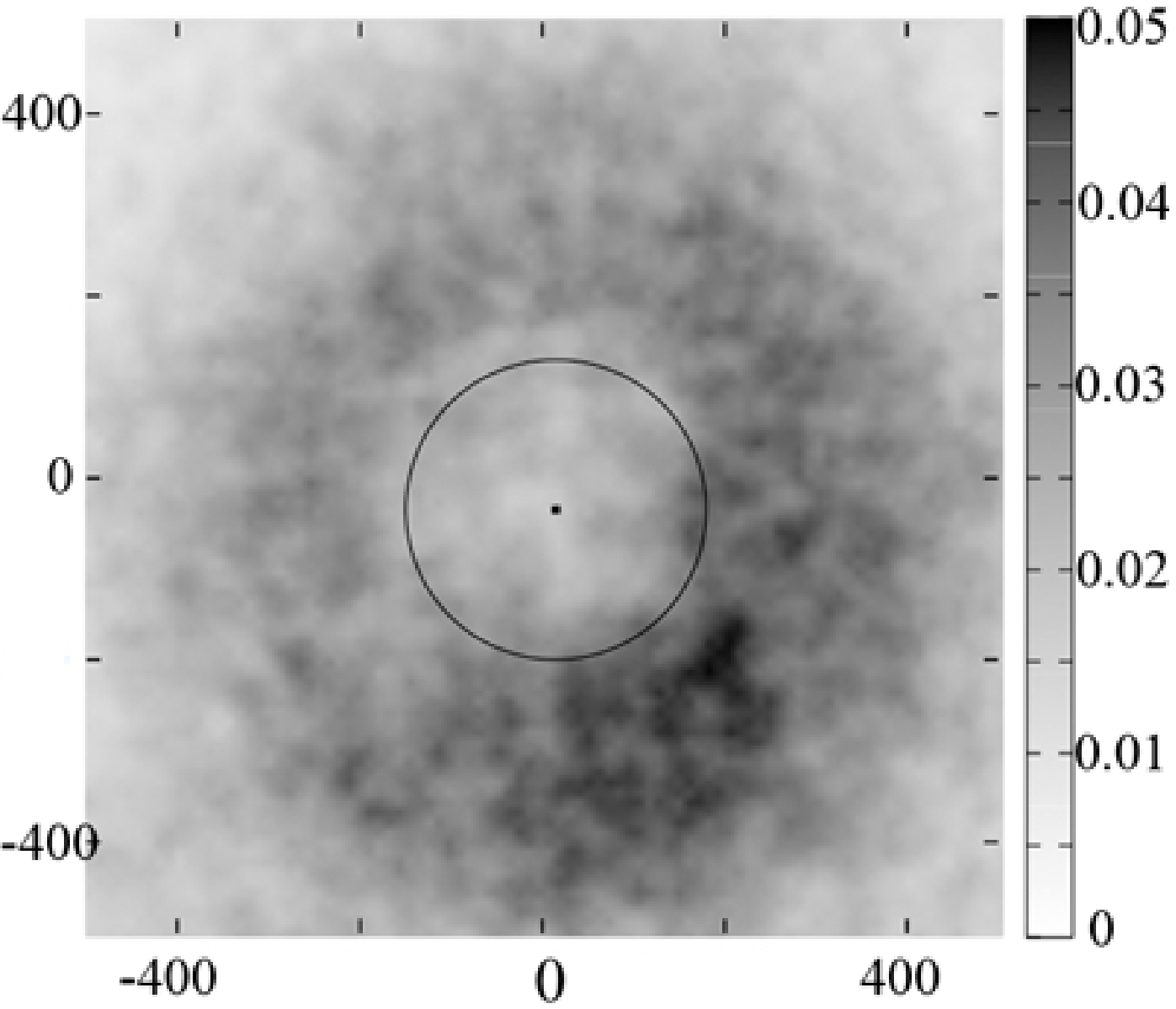}
}
\hspace{0.0cm}
\subfigure[VLT, H$\beta$(4861)] 
{
    \label{fig:sub:d}
    \includegraphics[width=5.5cm]{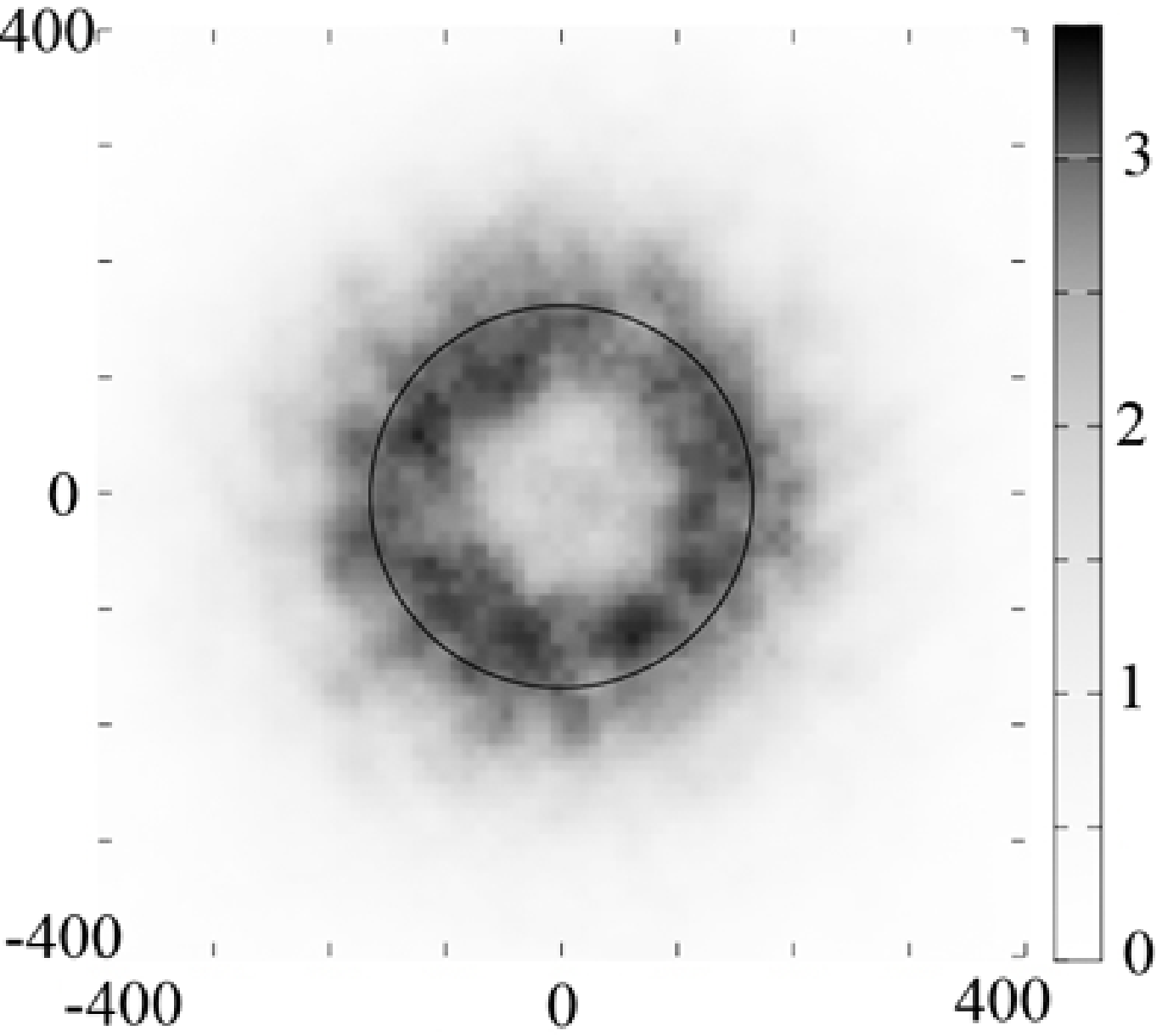}
}
\caption{Doppler tomograms for the most prominent lines. The center of
  symmetry in each map is marked as a dot and the outer disc
  radius is plotted as a circle. Both axis are in km\,s$^{-1}$.} 
\label{fig:tom} 
\end{figure*}

Doppler tomography~\citep{b33}, is an indirect imaging
method where the emission-line profiles, depending on phase, are
plotted onto a velocity scale. The method was used to visualize, in
velocity space, the origin of the emission features found in T Pyx. In
 Figure~\ref{fig:tom}, Doppler tomograms for the most prominent lines are
presented. For the reconstruction of the tomograms, the systemic velocity, $\gamma$ was set to zero (see Section~\ref{SV}). All lines show asymmetrically distributed emission from the disc. These asymmetric features are also seen in the trailed spectrograms (Figure~\ref{fig:spec}). No emission can be connected to the bright spot, the accretion stream or the donor star. The map of the Bowen blend was constructed from a composite of the three lines, NIII at $\lambda$4640.64 and CIII
at $\lambda$4650.1 and $\lambda$4647.4.


\subsection{A Radial-Velocity Study of T Pyx}
\label{RV}

\subsubsection{The Systemic Velocity}
\label{SV}

We noticed early on in our analysis for both the VLT and Magellan data
sets that the systemic velocity, $\gamma$, appeared to be varying with 
time. In order to confirm the reality of these variations, we
extracted spectra from a few VLT/GIRAFFE/FLAMES fibres that were
centered on knots in the nova shell surrounding T Pyx. The intrinsic
radial velocities of these large-scale knots are not expected to change
appreciably over a time-scale of months, so they provide a useful check
on the stability of our wavelength calibration. We found that the
shifts seen in the radial velocity data for the central object are
much larger than those seen for the 
knots, suggesting that these shifts cannot be explained by
instrumental effects alone. However, since we have only five nights
worth of VLT data  spread over two months, our data set is too sparse
to allow a careful study of trends in the apparent $\gamma$
velocities. For the purpose of the present study, we thus subtracted
the mean nightly $\gamma$ from all radial velocities before carrying
out any analysis. This should minimize the risk of biasing our
results. 

\subsubsection{The Velocity Semi-Amplitude}

\begin{figure}
\includegraphics[width=9cm]{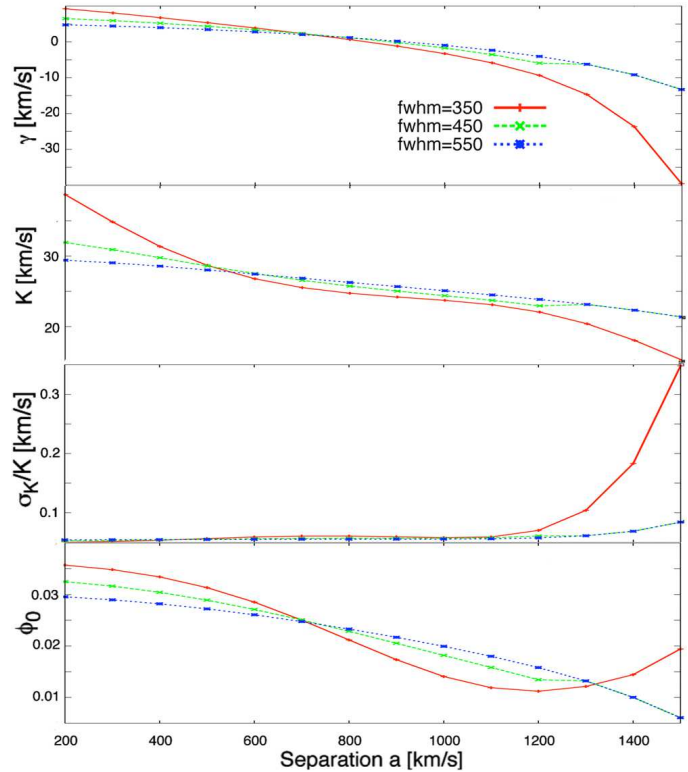}
\caption{Diagnostic diagram for the HeII line, Magellan. None of the
  parameters ($\gamma$, $K$ and $\phi_{0}$) are converging for any
  combination of the FWHM and separation.} 
 \label{fig:dd} 
\end{figure}

In order to establish the velocity semi-amplitude, $K1$, of the WD,
further analysis was pursued on the radial-velocity data obtained from
the double-Gaussian method. A least squares fit was made to the radial
velocity curves by keeping the orbital period fixed, but allowing the
three parameters, the systemic velocity, $\gamma$, the velocity
semi-amplitude, $K$, and the phase, $\phi_{0}$, to vary. Phase zero
corresponds to the photometric phase minimum light. The errors on the input radial velocity data,
provided in MOLLY, were used to calculate the $\chi^{2}$ and then
rescaled so that $\chi^{2} = 1$. The technique was applied to several
strong lines in both the VLT and Magellan data sets, but the radial
velocity data measured from the HeII line at 4686 \AA, and in
particular from the Magellan dataset, gave the most reliable fits. The
results were plotted in a diagnostic diagram (see~\citealp{b21};~\citealp{b25} and~\citealp{b6}). However, as can be seen in Figure~\ref{fig:dd}, the key parameters ($\gamma$, $K$ and $\phi_{0}$) did not converge convincingly for any combination of the FWHM and
separation. Normally, it is thought that the bright spot or other asymmetries in the disc are responsible for distorting the diagnostic diagram. However, in T Pyx the disc appears to be rather symmetric, and no contribution from the bright spot is seen (see Figure~\ref{fig:tom}). 
 We conclude that the radial velocity curves obtained from the
double-Gaussian method are, most likely, not tracking the velocity of
the primary WD, and therefore, $K1$, and $\phi_{0}$ cannot be
estimated reliably using this method.



\begin{figure}
\includegraphics[width=9cm]{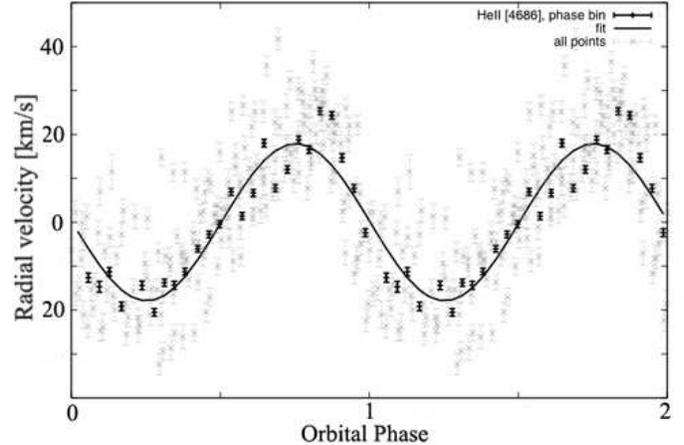}
\caption{Radial velocity curve of the HeII line, Magellan. Data in
  black is phase binned while all data are shown in grey. The best sin
  fit is over-plotted onto the phase-binned data.} 
\label{fig:rv} 
\end{figure}

\begin{figure*}
\includegraphics[width=18cm]{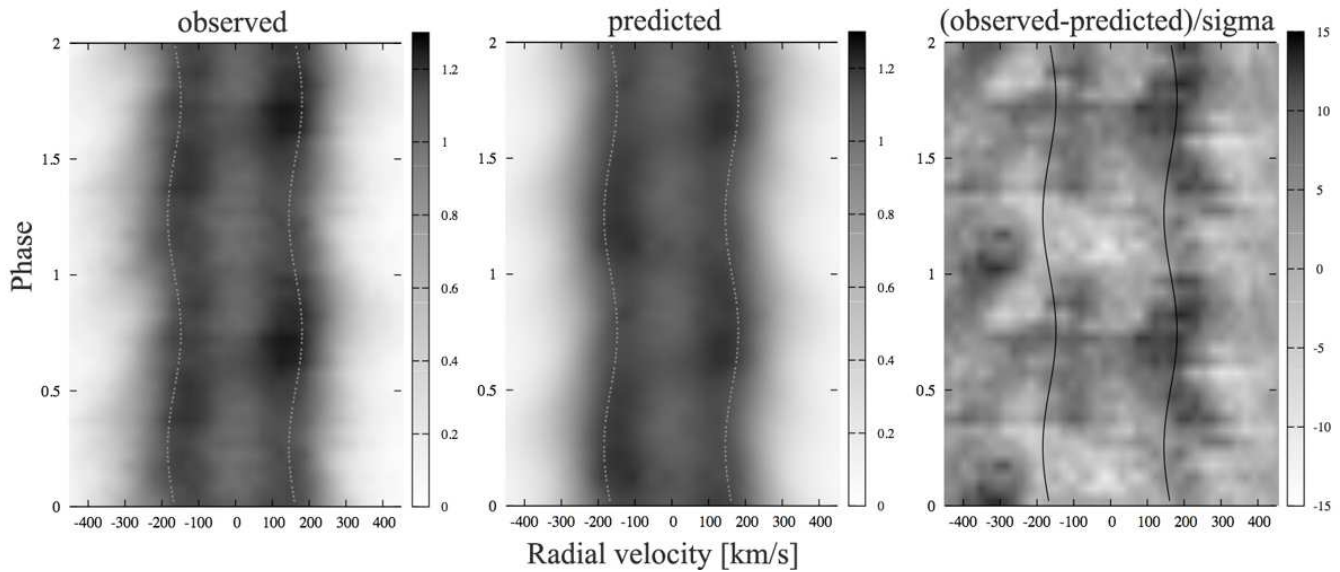}
\caption{Trailed spectrograms for the HeII line, Magellan.}
\label{fig:trail_heII} 
\end{figure*}

\noindent In order to improve on the results obtained from the
double-Gaussian method, we used two other techniques to obtain an
estimate of $K1$. First, we measured the velocity center of the lines
by fitting Gaussians to the individual spectral lines. Two Gaussians
were fitted to the double-peaked lines, keeping the FWHM and the rest
wavelength of the line fixed, but allowing the peak offsets and the
peak strengths to vary. Radial velocity curves were then plotted for
both blue and red wings. As can be seen from the trailed spectrograms,
Figure~\ref{fig:spec}, the red wing (in particular for the HeII
lines) does not show the same smooth orbital signal as the blue
wing. Consequently, only the data from the blue wing was taken into
account when measuring $K1$.  Several lines from both the VLT and
Magellan data sets were analyzed. The data were phase-binned to the
orbital resolution, velocity-binned in a $\sim$ 100 pixel wide region around
the line, and the nightly mean $\gamma$ was subtracted. HeII at
$\lambda$4686 is the strongest line and was ultimately used to
estimate our best-bet value of the velocity semi-amplitude of the WD:
$K1 = 17.9 \pm 1.6$ km\,s$^{-1}$. Figure~\ref{fig:rv} shows the
radial velocities obtained for the HeII line in the Magellan data
set. Data presented in black are phase-binned to the orbital
resolution afforded by our time-resolved spectroscopy, while all
invidual data points are shown in grey. Over-plotted onto the radial
velocity curve is the best sinusoidal fit, indicating a phasing close
to zero, $\phi_{0} = -0.03 \pm 0.03$. Figure~\ref{fig:trail_heII}
shows the final trailed spectrograms from the HeII at $\lambda$4686 in
the Magellan data set that was used to determine $K1$.

Second, $K1$ can also be measured from the Doppler
tomograms, by locating the center of symmetry of the disk emission
that dominates the maps. This center should be located at $V_{x} = 0$
and $V_{y}= - K1$ (eg.~\citealp{b24}). After masking out any
regions of strong asymmetric non-disk emission, the center of symmetry
was found by taking a trial point, smoothing the map azimuthally about
this point, and then subtracting this smoothed map from the real
map. If the trial point is far from the center of symmetry, the
resulting difference map will contain large asymmetric residuals. The
optimal center of symmetry is thus the trial point for which the
difference map exhibits the smallest residuals (as measured by the
standard deviation of the difference map). This procedure yielded
estimates for $K1$ that were consistent with those obtained from the
Gaussian fits ($14 < K1 < 19$ km\,s$^{-1}$). We thus adopt $K1 = 17.9 \pm 1.6$ km$\,s^{-1}$ as our best-bet estimate.

In closing this section, it should be acknowledged that our final
estimate of $K1$ is subject to considerably systematic
uncertainties. In particular, it is difficult to rule out that
whatever is preventing the double Gaussian method from converging may
also bias the Gaussian fits to the line peaks and the center of
symmetry of the tomograms. Strictly speaking, our estimate of $K1$
should thus perhaps be viewed as a lower limit. However, we are
reasonably confident that our measurements do trace the orbital motion
of the WD in T Pyx. This is mainly because the presence of persistent
double-peaked lines and ring-like structure in the tomograms implies
that there is a strong accretion disk component to the line
emission. This disk component must ultimately trace the motion of the
WD, and our two preferred methods are designed to exploit this in as
direct a fashion as possible, while simultaneously minimizing
contamination from asymmetrically placed emission regions.


\subsubsection{The Velocity at the Outer Disc Radius}

The peak-to-peak separations of the double peaked emission lines can
be used to estimate the projected velocity at the outer disc radius,
$v(R_{disk})\sin i$. We thus fitted Gaussians to the phase-binned, double-peaked H$\beta$, HeI and HeII lines in the Magellan data set. The peak-to-peak separations varies significantly with line but the mean value is close to that for HeI. We therefore use the HeI peak-to-peak separation for our estimation of $v(R_{disk})\sin i$, and adopt half the full range as a rough estimate of the associated error, $\Delta V_{peak-to-peak} = 290\,\pm\,26$ km\,s$^{-1}$. ~\cite{b23} investigated the exact relationship between the true $v(R_{disk})\sin i$ and the measured $\Delta
V_{peak-to-peak}$, taking into account the effects of instrumental
broadenings and different disk emissivity distributions. He found
that, for a wide range of parameters, $v(R_{disk})\sin i = 0.5 \Delta
V_{peak-to-peak}  / u$, where $u=1.05 \,\pm\, 0.05$. Taking this into
account, we find $v(R_{disk})\sin i = 138 \pm 15$ km\,s$^{-1}$.


\begin{figure}
\includegraphics[width=9cm]{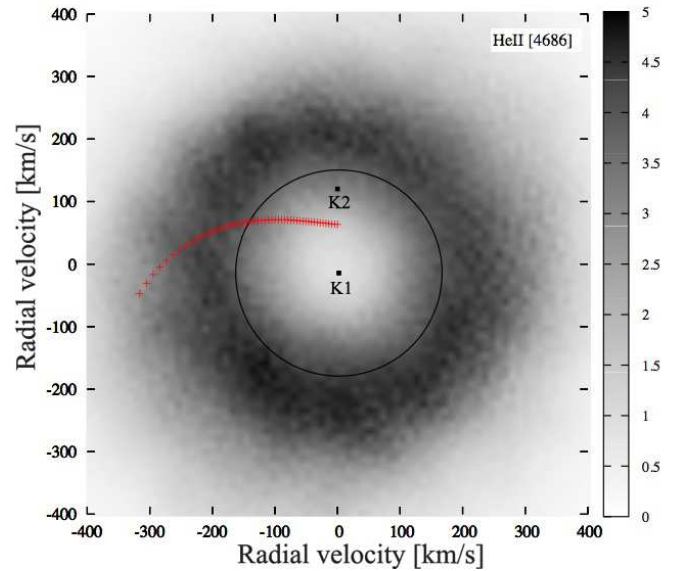}
\caption{Doppler tomogram for the HeII line, Magellan. The location of the WD, $K1$, the outer disc radius as well as the location of the donor star, $K2$, and the accretion stream is plotted onto the map.}
\label{fig:dopp_heII} 
\end{figure}


\section{Spectroscopic System Parameters}

If we are willing to assume that our estimates of $K1$ and $v(R_{disk})\sin i$ are valid, we can use our data to estimate several other system parameters. 

\subsection{Mass Ratio}

The method we use to estimate the mass ratio, $q$, is similar to that
described in~\cite{b27}. The basic idea is that, subject to fairly
benign assumption, the ratio $v(R_{disk})\sin i/K_1$ should be a
function of only $q$. More specifically, we assume that the disk is
circular, Keplerian, and, since T Pyx is a high-$\dot{M}$\ system,
that it extends all the way to the tidal radius. The disk radius can
then be taken as $R_{disk}/a = 0.60/(1+q)$~\citep{b28}. These assumptions, together with Kepler's third law, yield the expected relationship

\begin{equation}
\frac{v(R_{disk})\sin i}{K1}= 1.2909 \frac{1+q}{q}
\end{equation} 

\noindent Using this expression, we estimate a mass ratio of $q = 0.20 \pm 0.03$.
\\\\

Given a mass ratio and disk radius, it is possible to estimate the
expected phasing of the bright spot in the system, if we make the
usual assumption that the bright spot lies at the point where the
ballistic accretion stream from the L$_1$ point meets the disk
edge (see, for example,~\citealp{b34};~\citealp{b36}). Carrying out this calculation for T Pyx
suggests that the bright spot could be located at orbital phase $\phi
\simeq -0.03$. This is interesting, because it is consistent with the 
negligible phase offset we have found between photometric minimum
light and the red-to-blue crossing of the WD radial velocity
curve. Thus, based on phasing alone, the bright spot could be the
source of the photometric modulation. However, there are also other
interpretations for the photometric signature. For example, the
orbital signal could be dominated by a reflection effect, i.e. the
changing projected area of the irradiated face of the donor as it
moves around the orbit. This would explain why inferior conjunction of
the secondary corresponds to {\em minimum} light. On the other hand, 
it raises the question why we do not see the signature of the irradiated
donor in our spectroscopy. As noted above, there is no sign of the
narrow Bowen fluorescence lines that one might expect in the case of
strong donor irradiation. 

\subsection{Component Masses}

Recurrent novae are expected to harbor a more massive WD than most
ordinary CVs, and even than most classical novae. In a classical nova,
the mass of the WD is typically close to 1 M$_{\odot}$, both theoretically and 
observationally. Based on this, both~\cite{b20} and ~\cite{b18} agree
on a plausible mass range for the primary WD in T Pyx of 1.25 $-$ 1.4 M$_{\odot}$ (using
theoretical models by~\citealp{b31}). From the theoretical point
of view, a high-mass WD is needed to achieve a sufficiently short
outburst recurrence timescale.

In T Pyx, the bright accretion disc outshines any spectral signature
from the donor star, and we can therefore not constrain the donor mass
spectroscopically. However, we can use the period-density relation for
Roche-lobe filling secondary stars~\citep{b2} to set some
constraints. 

As shown by~\cite{b35} and~\cite{b8}, the donor stars
in ordinary CVs below the period gap are inflated by approximately 10\%
due to mass loss, relative to ordinary main sequence stars of the same
mass. Given T Pyx's peculiar evolutionary state, it is not clear if
this level of inflation is appropriate for T Pyx, and we therefore
adopt a conservative range of 0\% $-$ 20\% inflation. In order to
estimate the mass of the donor star, we thus take the 
theoretical main-sequence mass-radius relation from the 5 
Gyr isochrone of~\cite{b1}, adjust the stellar radius to
account for inflation, and then find the secondary mass that yields
the correct density for T Pyx's orbital period. This yields $M_{2} =
0.14 \pm 0.03$ M$_{\odot}$. Note that the inferred donor mass decreases
with increasing levels of radius inflation. The corresponding mass of
the WD is then $M_{1} = 0.7 \pm 0.2\, $M$_{\odot}$. 

Since this estimate of the WD mass is lower than expected for a
recurrent nova, we can also turn the problem around. If the mass of
the WD is $ > 1 $ M$_{\odot}$, as theoretical nova models imply~\citep{b31},
the mass of the donor becomes  > $0.2 \,$M$_{\odot}$, if our
estimate of the mass ratio is correct. 

All of these estimates should, of course, be taken with a grain of
salt. In particular, they rely on (i) the correctness of our measured
$K1$ and (ii) the assumption that the disk extends all the way to the
tidal radius. Despite these uncertainties, our calculations highlight
an important point: it is highly unlikely that the donor star in T Pyx
is already a brown dwarf. For example, if we retain the assumption of
maximal disk size,  $K1 \leq 7$~km\,s$^{-1}$ would be required in order for
$M_2 \leq 0.07$~M$_{\odot}$ for any $M_1 > 1~$M$_{\odot}$. We thus rule out
the possibility that T Pyx is a period bouncer in the usual sense,
i.e. that it had already reached the minimum period for ordinary CVs
($\simeq 70$~min theoretically, or $\simeq 80$~min observationally),
in which case its secondary would now be well below the
Hydrogen-burning limit.\footnote{T Pyx, is, of course, a ``period
bouncer'' in the basic sense that its orbital period is currently
increasing.}



\subsection{The Orbital Inclination}

The orbital inclination for T Pyx is thought to be low due to the
sharp spectral profiles and low radial velocity amplitude.~\cite{b22} suggested a lower limit of the orbital inclination for T Pyx of
$i \sim 6^{\circ}$ based on the peak-to-peak separation of the
H$\alpha$ line in their spectra.~\cite{b13} estimated an
inclination of $i \sim 10^{\circ} - 20^{\circ}$ due to the low amplitude of the
orbital signal, and~\cite{b20} estimated $i \sim 20^{\circ} - 30^{\circ}$.  
 
The inclination, $i$, can be estimated from our data via $K1 =
v_{1}\sin i$, where $v_{1} = (2 \pi a)/P \times (q/(q+1))$ and $a$ is the distance
between the two components obtained from Kepler's III law. This
provides a constraint on the system inclination of $i = 10 \pm 2$
degrees, for any reasonable combination of the component masses. We
stress that the real uncertainty on the inclination is bound to be
larger, because the formal error does not include systematic
uncertainties associated with, for example, possible bias in our $K1$
measurement and the assumption of tidal truncation in our derivation
of $q$.

\section{Summary and Discussion}  \label{SD}

The main result of our study is the spectroscopic determination of T
Pyx's orbital period, $P_{orb} \simeq 1.83$~hrs. This confirms that
the system is a CV below the period gap and implies that its current
accretion rate is at least 2 orders of magnitude higher than that of
an ordinary CV at this period. We also find that our spectroscopic
orbital period is consistent with the photometric ephemeris found for
T Pyx (\citealp{b13}, an updated version is given in
Section~\ref{PE}). This means not only that photometric timings can be
used as a more precise and convenient tracer of the orbital motion,
but also that the large period derivative required by the photometric
ephemeris marks a genuine change in the orbital period of the system.
In fact, the spectroscopic data are consistent with the photometric
period only if the period derivative is accounted for. The period
derivative obtained here from the latest combined photometric data
($\dot{P} = 6.7 \times 10^{-10}$) is slightly lower than that obtained
by~\cite{b13} from data up to 1997 ($\dot{P} \simeq 9 \times
10^{-10}$). A decrease in the rate of period change would be in line with~\cite{b18} scenario
that T Pyx's days as a high-$\dot{M}$ recurrent nova are numbered (at
least until its next ordinary nova eruption). In any case, the average
time-scale for period change found across all the photometric data are about $3 \times 10^5$~yrs.

We have also used our spectroscopic data to obtain estimates of other
key system parameters, most notably the radial velocity semi-amplitude
of the WD, $K1 = 17.9 \pm 1.6$ km\,s$^{-1}$, and the mass ratio, $q = 0.20 \pm
0.03$. The latter estimate rests on three key assumptions: first, that our
determination of $K1$ is correct, second, that our estimate
$v(R_{disk})\sin i$ is correct, and, third, that the accretion disk
around the WD in T Pyx is tidally limited. Taken at face value, this
relatively high value of the mass ratio implies that the donor star in
the system is not a brown dwarf. Thus T Pyx is not a period
bouncer. If we assume that the radius of the secondary is $10\% \pm
10\%$ inflated relative to an ordinary main sequence star of the same
mass, we find that its most likely mass is $M_{2} = 0.14 \pm 0.03$
M$_{\odot}$.

Overall, the physical picture that emerges from our study is
consistent with the scenario proposed by~\cite{b18}. In particular,
they suggest that, prior to the 1866 eruption, T Pyx was an ordinary
CV. That eruption then triggered a  high-$\dot{M}$ wind-driven phase,
as suggested by~\cite{b7} to account for T Pyx's exceptional
luminosity. However, this phase is not quite self-sustaining, so that
T Pyx is now fading and perhaps not even a RN anymore. In line with
this picture, we find that the mass ratio and donor mass we derive are
not abnormally low for a CV at its orbital period. This shows that the
present phase of high-$\dot{M}$ accretion cannot have gone on for too
long already. The hint of a declining period derivative may point in
the same direction, but this needs to be confirmed. 

Does all this mean that the phase of extraordinarily high accretion
rates T Pyx is currently experiencing will have no lasting impact on
its evolution? Not necessarily. In a stationary Roche-lobe-filling
system, the orbital period derivative and total mass-loss rate from
the donor (wind loss + mass transfer) are related via
\begin{equation}
\frac{\dot{P}_{orb}}{P_{orb}} = \frac{3 \zeta -1}{2}
\frac{\dot{M}_2}{M_2}
\end{equation}
where $\zeta$ is the donor's mass-radius index. In T Pyx, which has an
increasing orbital period, we will take $\zeta \simeq -1/3$, which is
appropriate for adiabatic mass-loss from a fully convective star
(e.g.~\citealp{b7}). We thus expect that $\dot{M_2}/M_2 \simeq
{\dot{P}_{orb}}/{P_{orb}}$, which suggests a typical mass-loss rate
from the donor of $\dot{M_2} \sim 5\times10^{-7}$~M$_{\odot}$/yr in its current
state. (The accretion rate onto the WD can be lower than this, since,
in the wind-driven scenario, much of this mass escapes in the form of
an irradiation-driven outflow from the donor.)  If every ordinary nova
eruption in T Pyx is followed by $\sim$ 100 yrs of such
high-$\dot{M}_2$ evolution, the total mass loss from the donor in the
luminous phase is $\sim 5 \times 10^{-5}$ M$_{\odot}$ between any two
such eruptions. This needs to be compared to the mass lost from the
donor during the remaining part of the cycle. If this is driven by
gravitational radiation, the mass loss rate from the donor will be
$\dot{M}_2 \simeq 5 \times 10^{-11}$~M$_{\odot}$/yr~\citep{b7}. The
recurrence time of ordinary nova eruptions for such a system is on the order of $10^5$ yrs~\citep{b31}, so the total
mass lost from the donor during its normal evolution (outside the
wind-driven phase) is $5 \times 10^{-6}$ M$_{\odot}$. {\em This shows
that the long-term secular evolution may be dominated by its
high-$\dot{M}$ wind-driven phases, even if the duty cycle of these
phases is very low (e.g. 0.1\% for the numbers adopted above).}

It is finally tempting to speculate on the relevance of ``T Pyx-like''
evolution for ordinary CVs. At first sight, the numbers above suggest
that the evolution of a CV caught in such a state may be accelerated
by about an order of magnitude. This is interesting, since it could
bear on the long-standing problem that there are fewer short-period
CVs and period bouncers in current samples than theoretically expected
(e.g.~\citealp{b13};~\citealp{b14};~\citealp{b15}; Pretorius \& Knigge
2008ab). Moreover, this channel need not be limited to systems containing high-mass WDs that are capable of
becoming RNe. After all, it is not the recurrent nova outbursts that
are of interest from an evolutionary point of view, but simply the
existence of a prolonged high-$\dot{M}$ phase in the aftermath of nova
eruptions. In CVs containing lower-mass WDs, such a phase may still
occur, although its evolutionary significance could still depend on
the WD mass. For example, the interoutburst time-scale is longer for
low-mass WDs, and the duration of the high-$\dot{M}$ phase could also
scale with WD mass. One obvious objection to this idea is that, observationally, most nova
eruptions are not followed by centuries- (or even decades-) long
high-$\dot{M}$ phases. However, this need not be a serious issue. Most
observed novae are long-period systems, so if the triggering of a
wind-driven phase requires a fully convective donor, most novae would
not be expected to enter such a phase. It may be relevant in this
context that at least one other short-period nova -- GQ Mus --
exhibited an exceptionally long post-outburst supersoft phase of $\sim
10$~years~\citep{b12}.~\cite{b3} have also
suggested that the duration of the supersoft phase in novae may scale
inversely with orbital period. 

However, there is another important consequence to the idea that the
evolution of many short-period CVs is accelerated by ``T Pyx-like''
high-luminosity phases. The orbital period-derivative is positive in
the high-$\dot{M}$ phase, but negative during the remaining times of
GR-driven evolution. But since $\dot{P}_{orb}/P_{orb} \simeq
\dot{M}_2/M_2$ in {\em both} phases, the sign of the secular
(long-term-average) period derivative will generally correspond to the
phase that dominates the secular evolution. So whenever wind-driving
dramatically accelerates the binary evolution, the direction of period
evolution will also be reversed. Is this a problem? Perhaps. The
recent detection of the long-sought {\em period spike} in the
distribution of CV orbital periods~\citep{b4} suggests
that there is, in fact, a reasonably well-defined minimum period for
CVs, as has long been predicted by the standard model of CV evolution
(e.g.~\citealp{b9}). On the other hand, the location of the observed
spike ($\simeq 83$~mins) is significantly different from the expected
one ($\simeq 65-70$~min). Is it possible that the observed period
minimum corresponds to the onset of wind-driving in most CVs? T Pyx,
with $P_{orb} \simeq 110$~min would then have to be an outlier,
however, perhaps because of an unusually high WD mass.

The idea that T Pyx-like phases may significantly affect the
evolution of many CVs is, of course, highly speculative, and we do 
not mean to endorse it too strongly. However, it highlights the
importance of understanding T Pyx: until we know what
triggered the current high-luminosity state, it will remain difficult 
to assess the broader evolutionary significance of this phase. Note
that the apparent uniqueness of T Pyx is not a strong argument against
such significance. For example, if the duty cycle of high-luminosity
phases is $\sim 0.1$\%, as suggested by the numbers above, we should
not expect to catch many CVs in this state. Thus T Pyx could be the
tip of the proverbial iceberg.

\section*{Acknowledgments}

We thank Joe Patterson for valuable comments and for kindly providing
us with the a compilation of recent photometric timings for T Pyx. We also thank Gijs Roelofs for obtaining part of the data. Danny Steeghs acknowledges a STFC Advanced Fellowship.

\bsp

\label{lastpage}

\end{document}